\documentclass[12pt]{article}
\usepackage{graphicx}
\usepackage{amsmath}
\usepackage{cite}
\usepackage{verbatim}
\usepackage{url}
\usepackage{hyperref}

\title{Detecting and Mitigating SQL Injection Vulnerabilities in Web Applications}
\author{Sagar Neupane \\ University of West London, Ealing, United Kingdom}

\begin{document}

\maketitle

\begin{abstract}
SQL injection (SQLi) remains a critical vulnerability in web applications, enabling attackers to manipulate databases through malicious inputs. Despite advancements in mitigation techniques, the evolving complexity of web applications and attack strategies continues to pose significant risks. This paper presents a comprehensive penetration testing methodology to identify, exploit, and mitigate SQLi vulnerabilities in a PHP-MySQL-based web application. Utilizing tools such as OWASP ZAP, sqlmap, and Nmap, I demonstrate a systematic approach to vulnerability assessment and remediation. My findings underscore the efficacy of input sanitization and prepared statements in mitigating SQLi risks, while highlighting the need for ongoing security assessments to address emerging threats. This study contributes to the field by providing practical insights into effective detection and prevention strategies, supported by a real-world case study. The complete source code and datasets used in this research, are hosted on GitHub. The repository can be accessed \href{https://bit.ly/4dQHswx}{here}.
\end{abstract}

\section{Introduction}

SQL injection (SQLi) is a pervasive security vulnerability that allows attackers to inject malicious SQL code into database queries, potentially compromising sensitive data, bypassing authentication, or disrupting services (Hasan et al., 2019). Despite being a well-documented issue, SQLi remains prevalent due to inadequate input validation, poor coding practices, and the increasing sophistication of attack techniques (Nasereddin et al., 2021). According to  MITRE Corporation's 2021 CWE definition, SQLi (CWE-89) is a critical software weakness, underscoring its severe impact (MITRE Corporation, 2021).

Organisations have been severely harmed by recent SQL Injection attacks. For instance, Tang et al. (2020) report that a SQL injection attack on the National Job Portal revealed the personal information of around thirty million Indian job searchers. The British Airways website was hacked in 2019 via SQL injection, Kumar et al. (2022) claim, potentially putting the personal and financial information of hundreds of thousands of consumers at risk. According to Farah et al. (2016), a cyberattack in 2016 that employed SQL injection to infiltrate the Bangladesh Central Bank's systems cost the bank 81 million dollars.

These recent SQL injection attacks have functioned as a reminder for businesses to give their websites and databases top priority when it comes to security. Implementing frequent security assessments and penetration testing is essential to finding vulnerabilities and patching them before hackers may take advantage of them. Attacks using SQL injection can have serious repercussions, such as monetary loss, reputational damage, and legal responsibilities. The detection and prevention of SQL injection attacks is a difficult undertaking for website owners, particularly if the website is large and complicated. Additionally, patching a vulnerability may not always be simple, and a patch that is not thorough can leave a website open to assaults.

This research conducts an in-depth security assessment of a PHP-MySQL-based web application, focusing on SQLi vulnerabilities. Through penetration testing, I aimed to identify exploitable weaknesses, evaluate their impact, and implement robust mitigation strategies. Tools such as OWASP ZAP (Nájera-Gutiérrez, 2016), sqlmap (Gunawan et al., 2018), and Nmap (Liao et al., 2020) are employed to simulate real-world attack scenarios, providing a practical framework for vulnerability detection and remediation. The study also explores the effectiveness of preventive measures like prepared statements and input sanitization, drawing on insights from prior research (Rahman et al., 2015; Alomari and Jerisat, 2021).

In this paper, penetration testing is used to evaluate a website's security with an emphasis on SQL injection attacks. An intricate SQL injection assault has been launched against the website, which was created using PHP and SQL. The website has been retested to guarantee its security once the attack's vulnerabilities are corrected. By creating a website with PHP and SQL, conducting a thorough SQL injection attack, and fixing the vulnerabilities found during the attack, this study addresses these challenges. To conduct the security evaluation, a thorough security assessment strategy was established.

\begin{figure}[h!]
    \centering
    \includegraphics[width=0.8\textwidth]{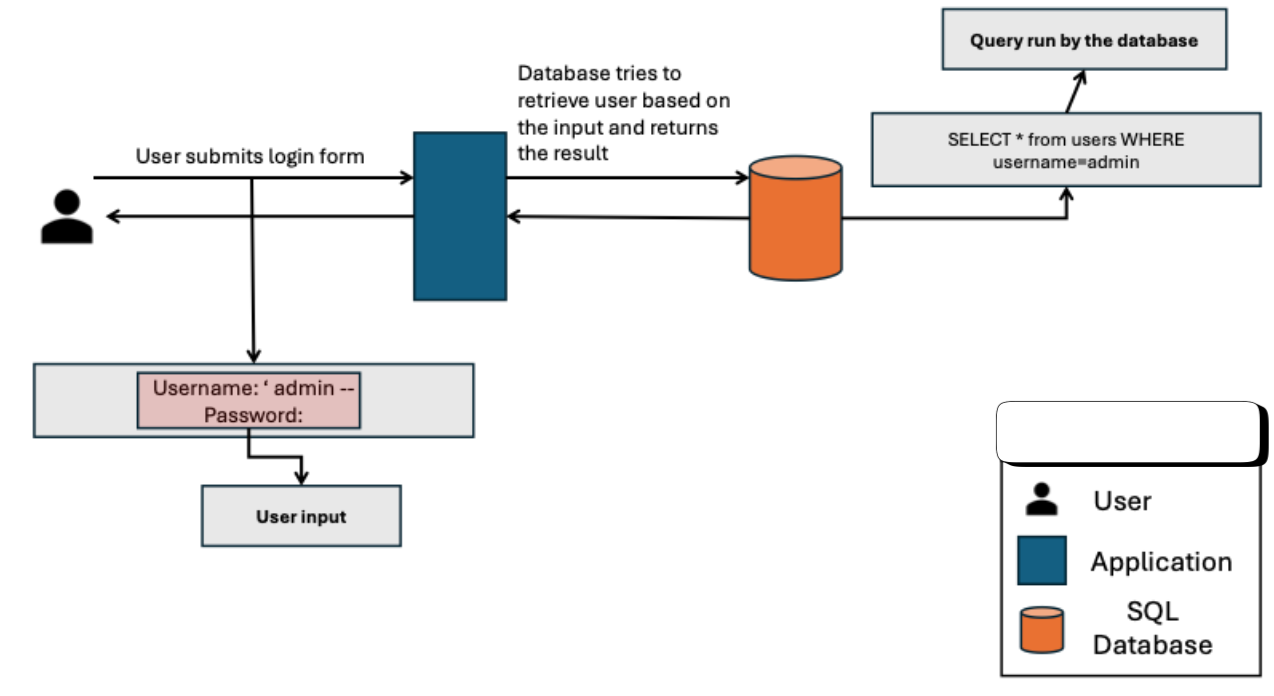}
    \caption{SQL injection attack flow (Hasan et al., 2019).}
    \label{fig:figure2.1}
\end{figure}

\section{Problem Statement}
 
SQL injection attacks remain a persistent and escalating threat to website security, despite the availability of best practices and mitigation techniques. These attacks exploit vulnerabilities in web applications, compromising sensitive data, disrupting operations, and causing significant financial and reputational damage. To address this ongoing challenge, there is a critical need to assess the effectiveness of existing security measures, develop innovative strategies, and foster practical skills in website security and penetration testing. The following challenges highlight the key barriers to preventing and mitigating SQL injection attacks effectively.

\begin{enumerate}
    \item \textbf{Limited Developer Awareness and Expertise}: Many website developers lack sufficient knowledge of SQL injection attacks, hindering their ability to implement secure coding practices and recognize vulnerabilities during development.
    \item \textbf{Ineffective or Outdated Security Practices}: Current security measures and best practices for preventing SQL injection attacks are often inadequate or outdated, leaving websites vulnerable to exploitation.
    \item \textbf{Increasing Sophistication of Cybercriminals}: Cybercriminals are employing advanced techniques to bypass even robust security measures, increasing the difficulty of defending against SQL injection attacks.
    \item \textbf{Resource Constraints for Small Organizations}: Small businesses and nonprofit organizations often lack the financial resources to hire skilled security professionals or invest in advanced security solutions, making them prime targets for SQL injection attacks.
    \item \textbf{Complexity of Modern Web Applications}: The growing complexity of web applications and databases introduces new vulnerabilities, requiring innovative strategies to secure systems against SQL injection attacks.
    \item \textbf{Lack of Standardization in Security Practices}: The absence of consensus and standardized guidelines for preventing SQL injection attacks within the web development community leads to inconsistent security implementations.
    \item \textbf{Emerging Threats from Cloud-Based Systems}: The increasing adoption of cloud-based web applications and databases introduces unique security challenges, necessitating new approaches to mitigate SQL injection risks.
    \item \textbf{Insufficient Penetration Testing and Validation}: Many organizations fail to conduct regular penetration testing or vulnerability assessments, leaving undetected SQL injection vulnerabilities in production environments.
    \item \textbf{Slow Adoption of Secure Development Frameworks}: The reluctance or inability to adopt secure development frameworks and tools that inherently mitigate SQL injection risks perpetuates vulnerabilities in web applications.
    \item \textbf{Inadequate Training and Skill Development}: There is a shortage of accessible, practical training programs focused on building hands-on skills in identifying, preventing, and mitigating SQL injection attacks among developers and security professionals.
\end{enumerate}

\section{Aims}

This research aims to advance the understanding and mitigation of SQL injection vulnerabilities in web applications through a rigorous, evidence-based evaluation of security mechanisms and penetration testing methodologies. It seeks to systematically assess the susceptibility of a PHP and SQL-based web application to SQL injection attacks, identifying critical vulnerabilities and their exploitation vectors. The study will critically evaluate the effectiveness of contemporary security measures and best practices, focusing on their applicability to complex web architectures, and formulate evidence-based recommendations to enhance web application security, addressing emerging SQL injection threats. Additionally, it aims to contribute novel insights to the academic and professional discourse on web security, enriching the existing body of knowledge on SQL injection countermeasures, while fostering advanced, transferable expertise in penetration testing and secure software development among researchers.

\section{Objectives}

    \begin{enumerate}
        \item Design and implement a controlled PHP and SQL-based web application with intentional vulnerabilities to serve as a testbed for SQL injection penetration testing, ensuring replicability and scalability for research purposes.
        \item Conduct comprehensive penetration testing using advanced tools, such as SQLMap (Gunawan et al., 2018), Burp Suite (Wear, 2018), OWASP ZAP(Nájera-Gutiérrez, 2016), and Nmap (Liao et al., 2020), to identify and exploit SQL injection vulnerabilities, employing both automated and manual techniques.
        \item Cultivate specialized skills in web application security and penetration testing, mastering industry-standard tools and methodologies, including vulnerability scanning, exploit development, and network reconnaissance.
        \item Perform a thorough security audit of the testbed application, integrating multiple penetration testing approaches, such as black-box, gray-box, and white-box testing, to evaluate the attack surface and vulnerability impact (Hussain and Singh, 2015).
        \item Quantitatively and qualitatively assess the efficacy of security mechanisms, including input validation, parameterized queries, and web application firewalls (WAFs)  (Prandl et al., 2015), in mitigating SQL injection risks, using metrics like false positive rates and exploit success rates.
        \item Analyze the consequences of successful SQL injection attacks on the testbed application, quantifying impacts on data integrity, confidentiality, availability, and potential downstream effects, such as reputational harm and regulatory non-compliance.
        \item Develop a comprehensive set of recommendations for improving web application security, incorporating advanced strategies, such as secure coding frameworks and runtime application self-protection, while addressing challenges in cloud-based and distributed systems.
        \item Document and disseminate findings through high-impact academic publications and presentations, contributing to the global knowledge base on SQL injection prevention and web security best practices.
    \end{enumerate}
    
\section{Research Questions}

The persistent threat of SQL injection attacks continues to challenge the security of web applications, particularly in PHP and SQL-based systems, necessitating advanced research to address evolving vulnerabilities and attack vectors. These questions reflect the current needs of the cybersecurity landscape, emphasizing modern web architectures, emerging defense strategies, and standardized solutions to enhance resilience against sophisticated threats.

\begin{enumerate}
    \item What are the prevalence and impact of SQL injection attacks on modern web applications, particularly in PHP and SQL-based systems, and how do these vary across industries such as finance, healthcare, and e-commerce?
    \item What specific vulnerabilities in PHP and SQL codebases, including those in cloud-based and distributed architectures, are most commonly exploited by SQL injection attacks, and how do these vulnerabilities evolve with emerging web development frameworks?
    \item How effective are current security mechanisms, such as parameterized queries, input sanitization, and web application firewalls, in mitigating SQL injection attacks under diverse attack scenarios, including advanced persistent threats (Prandl et al., 2015)?
    \item What novel or hybrid defense strategies, incorporating secure coding practices, runtime application self-protection, and machine learning-based anomaly detection, can significantly enhance the prevention of SQL injection attacks in complex web applications?
    \item How can standardized, interoperable security protocols and frameworks be developed and adopted to protect PHP and SQL-based web applications from SQL injection vulnerabilities, particularly in resource-constrained environments like small organizations?
    \item What are the measurable impacts of successful SQL injection attacks on data integrity, confidentiality, and system availability, and how do these translate into broader consequences, such as regulatory non-compliance, financial losses, and reputational damage?
    \item How can automated penetration testing and vulnerability assessment tools be optimized to detect and prioritize SQL injection vulnerabilities in real-time, especially in large-scale, dynamic web applications?
    \item What role can advanced training programs and simulation-based learning play in equipping developers and security professionals with the skills to identify, mitigate, and prevent SQL injection attacks in modern web ecosystems?
\end{enumerate}
 
\section{Literature Review \& Gap Analysis}

In their 2019 study, Hasan et al. proposed a machine learning-based approach to detect SQL injection attacks, analyzing a dataset of 20,000 SQL queries (10,000 benign, 10,000 malicious) using classifiers such as Random Forest, achieving a 99.2\% accuracy rate, motivated by the need to bolster website security against prevalent SQLi threats; however, the limited dataset size and omission of network delay or packet loss effects constrained generalizability, which this paper addresses through real-world penetration testing with tools like OWASP ZAP and sqlmap, incorporating network constraint evaluations.

In a 2015 investigation, Appelt et al. evaluated the efficacy of firewalls in mitigating SQL injection attacks, employing automated and manual tactics to test network-based and application-based firewalls, finding the former more effective, driven by the need to assess firewall protection in web applications; the study’s exclusive focus on firewalls, neglecting other strategies and false positives, prompted this paper to assess multiple mitigation techniques, including input validation and parameterized queries, while addressing false positives and application complexity.

In their 2015 research, Rahman et al. developed an algorithm to detect SQL injection in e-commerce websites by comparing user input to predefined query patterns, achieving high accuracy with low false positives to safeguard sensitive data; its dependence on static patterns limited detection of novel attacks and ignored vectors like HTTP requests, leading this paper to propose a comprehensive detection approach for sophisticated SQLi patterns across multiple vectors via penetration testing.

In a 2020 study, Aliero et al. introduced SQLIA, a Python-based tool leveraging Scikit-learn to automate SQL injection vulnerability detection, achieving 98.2\% accuracy to counter data breaches; incomplete performance evaluation across diverse datasets and complex attacks led this paper to rigorously analyze SQLIA’s capabilities and test its effectiveness against advanced SQLi scenarios in various web applications.

In their 2019 work, Sarjitus and El-Yakub proposed modifying server-side code to detect SQL injection vulnerabilities, effectively identifying flaws in a test application to enhance web security; the requirement for code modifications and deep architectural knowledge prompted this paper to develop a universal detection method applicable without extensive code changes.

In a 2020 analysis, Tripathy et al. employed machine learning to detect SQL injection at the application layer by analyzing HTTP traffic features, surpassing signature-based methods to address bypassable detection techniques; reliance on HTTP traffic and preprocessing scalability issues led this paper to propose a scalable SQLi detection method minimizing preprocessing and HTTP dependency.

In their 2016 study, Charania and Vyas developed WAVES, a black-box tool to detect SQL injection vulnerabilities through HTTP request parsing, effectively identifying flaws to improve automated detection; its potential to miss vulnerabilities and lack of limitation discussion prompted this paper to critically evaluate detection tools and suggest improvements for comprehensive vulnerability coverage.

In a 2019 investigation, Zhang developed an AI classifier to detect SQL injection vulnerabilities in PHP code, using supervised machine learning on a dataset of vulnerable and non-vulnerable PHP code snippets, outperforming static analysis techniques to address persistent web security issues; the study’s focus on PHP alone and lack of consideration for machine learning biases prompted this paper to examine biases in machine learning models and propose methods for detecting SQLi vulnerabilities across multiple programming languages.

In their 2018 study, Katole et al. proposed a boundary-based method to detect SQL injection attacks by extracting query boundaries and comparing original and modified SQL queries, achieving accurate detection even against obfuscated attacks, motivated by the need to reliably identify attacker modifications in web applications; the computationally intensive boundary extraction and potential for false positives led this paper to suggest a framework for integrating this method with existing security tools to enhance efficiency and reduce manual inspections.

In a 2020 study, Hlaing and Khaing developed a lexicon-based approach to detect SQL injection attacks by identifying query tokens using a dictionary of common SQLi phrases, achieving high accuracy with low false positives to address limitations in existing detection methods; the reliance on a predefined dictionary and lack of query context consideration, which may cause false positives, prompted this paper to propose a hybrid method combining lexicon-based and context-aware techniques for more robust SQLi detection.

In their 2019 research, Yip et al. proposed an adaptive learning-based SQL injection detection system using ensemble feature selection to classify web requests, achieving high detection rates with low false positives to counter sophisticated attacks; the lack of discussion on real-world implementation challenges, such as performance impacts, and focus solely on detection prompted this paper to evaluate practical implementation feasibility and explore preventive measures for SQLi attacks.

In a 2021 investigation, Alomari and Jerisat employed machine learning to detect SQL injection attacks, using feature selection algorithms and classifiers like Random Forest on data from sources like the National Vulnerability Database, achieving over 98\% accuracy to address new attack vectors; the study’s focus on detection without prevention strategies led this paper to propose preventive methods through penetration testing and suggest rapid implementation of machine learning-based detection.

In their 2021 study, Zhang et al. developed an intelligent SQL injection detection technique using machine learning with feature selection, testing six algorithms on a dataset of safe and malicious traffic, achieving 98.25\% accuracy to overcome limitations of signature-based detection; the lack of testing on a larger, more diverse dataset limited generalizability, which this paper addresses by evaluating the technique on varied datasets and exploring ethical considerations in penetration testing.

In a 2021 analysis, Al-Saleh and Saito proposed a machine learning-based framework combining static analysis and dynamic testing to detect SQL injection vulnerabilities, outperforming existing tools in precision and identifying new flaws to enhance web application testing; the framework’s failure to account for the evolving nature of web applications and sophisticated attacks prompted this paper to develop a more adaptive testing approach capable of detecting emerging SQLi threats.

In their 2020 study, Chen et al. developed a deep belief network with transfer learning to detect SQL injection vulnerabilities, achieving a 98.64\% detection rate with a low false positive rate to adapt to diverse web applications; the reliance on a small dataset for fine-tuning and lack of focus on feature extraction impacts led this paper to propose a more generalizable approach with enhanced feature selection techniques for broader applicability.

Several gaps existed in prior research on SQL injection detection and mitigation, including limited dataset sizes, neglect of real-world network constraints, reliance on static patterns, and insufficient evaluation of complex attack scenarios. This research addresses these shortcomings by conducting comprehensive penetration testing on a PHP and SQL-based web application, employing tools like OWASP ZAP (Nájera-Gutiérrez, 2016) and sqlmap (Gunawan et al., 2018) to simulate realistic attack conditions, and proposing adaptive, scalable mitigation strategies such as input validation, parameterized queries, and hybrid detection methods to enhance web application security against evolving SQLi threats.

\section{Methodology}

This study employs a systematic penetration testing methodology to assess SQLi vulnerabilities in a PHP-MySQL-based web application. The methodology comprises three phases: preparation, execution, and patching, as detailed below.

\subsection{Preparation}

The preparation phase established a structured foundation for an ethical and systematic security assessment, ensuring legal compliance and comprehensive reconnaissance. The scope was defined to target a locally developed e-commerce website, constructed with PHP and MySQL, hosted on a XAMPP server (Dvorski, 2007) running Apache and MySQL services. The assessment focused on front-end interfaces, including product detail, login, and checkout pages, and their back-end database interactions, with an emphasis on SQL injection vulnerabilities. Ethical considerations were addressed by securing explicit permission from the website’s developer and confining testing to a controlled environment to prevent impact on live systems. Reconnaissance utilized tools such as Nmap (Liao et al., 2020) and Whois (Velu and Beggs, 2019) to gather infrastructure details. Nmap was executed to identify open ports (e.g., 80 for HTTP, 3306 for MySQL), services. Whois lookup provided domain registration details, limited to localhost in this test environment. Relevant stakeholders, including the academic supervisor and peers, were formally notified to ensure transparency and alignment with research objectives. This phase facilitated a thorough understanding of the target system, enabling precise and authorized penetration testing.

\begin{figure}[h!]
    \centering
    \includegraphics[width=0.8\textwidth]{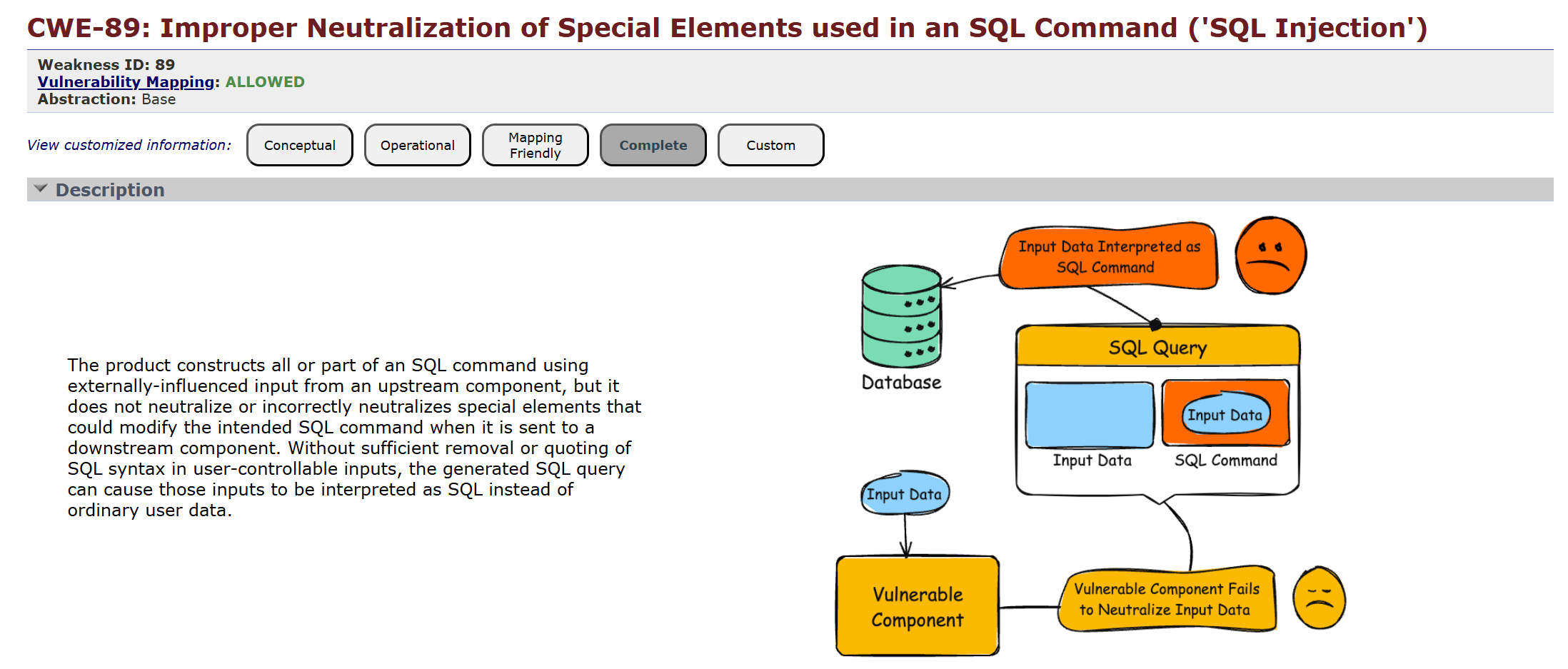}
    \caption{CWE-89 (SQL Injection) ( MITRE Corporation, 2021).}
    \label{fig:figure4.20}
\end{figure}

\subsection{Execution}

The execution phase involved active penetration testing to identify and exploit SQL injection vulnerabilities, employing a blend of automated and manual techniques aligned with industry standards. Testing targeted the e-commerce website’s vulnerable pages, particularly the product detail page (detail.php), which processed user inputs via GET parameters (e.g., pro\_id). Automated scanning was conducted using OWASP ZAP (Nájera-Gutiérrez, 2016), configured for a comprehensive scan to detect potential vulnerabilities, identifying a SQL injection flaw classified under CWE-89 ( MITRE Corporation, 2021). To validate this, sqlmap was utilized to perform targeted SQL injection attacks, enumerating databases, extracting table structures, and retrieving sensitive data such as admin credentials and customer details (Gunawan et al., 2018). Manual testing supplemented automation by injecting payloads such as OR 1 = 1 into input fields to observe database responses, confirming unauthorized data access. The approach began with a black-box perspective, transitioning to grey-box testing with limited code access to simulate realistic attack scenarios, ensuring robust vulnerability detection (Hussain and Singh, 2015).

\subsubsection{Vulnerabilities Identified}

The penetration testing conducted on the e-commerce website revealed critical SQL injection vulnerabilities that enabled unauthorized access to sensitive database contents, posing severe risks to data integrity, confidentiality, and organizational reputation. The primary vulnerability was identified in the website’s product detail page (detail.php), where user inputs via GET parameters (e.g., pro\_id) were processed without proper sanitization or parameterization, allowing malicious SQL queries to be injected into the back-end MySQL database(Grippa and Kuzmichev, 2021). Automated scanning with OWASP ZAP (Nájera-Gutiérrez, 2016) detected this flaw, classifying it under CWE-89 (SQL Injection) ( MITRE Corporation, 2021), indicating the potential for attackers to manipulate database queries. Subsequent validation using sqlmap (Gunawan et al., 2018) confirmed the vulnerability’s severity, as commands successfully enumerated the database schema, listing databases such as ecom, and extracted table structures and sensitive data, including administrator credentials (e.g., usernames and passwords from the admins table), customer personal information (e.g., names and email addresses from the customers table), and billing details (e.g., payment information from the billing\_details table). Manual testing further corroborated these findings by injecting payloads like ' OR 1=1 -- into input fields, which bypassed authentication mechanisms and retrieved unauthorized data, demonstrating the absence of input validation or query escaping mechanisms. 

The vulnerability stemmed from insecure coding practices, specifically the direct concatenation of user inputs into SQL queries, which facilitated arbitrary query execution (Neumann and Kemper, 2015). This flaw enabled potential data theft, as attackers could exfiltrate entire database contents, compromising user privacy and organizational assets. Additionally, the exposure of sensitive information risked loss of confidentiality, as unauthorized access to customer and admin data could lead to exploitation in phishing campaigns or identity theft. The analysis underscores that these vulnerabilities could precipitate reputational harm, as public disclosure of a data breach would erode customer trust, diminish platform engagement, and invite negative media scrutiny, potentially leading to loss of business and increased costs for remediation efforts. 
 
\begin{figure}[h!]
    \centering
    \includegraphics[width=0.8\textwidth]{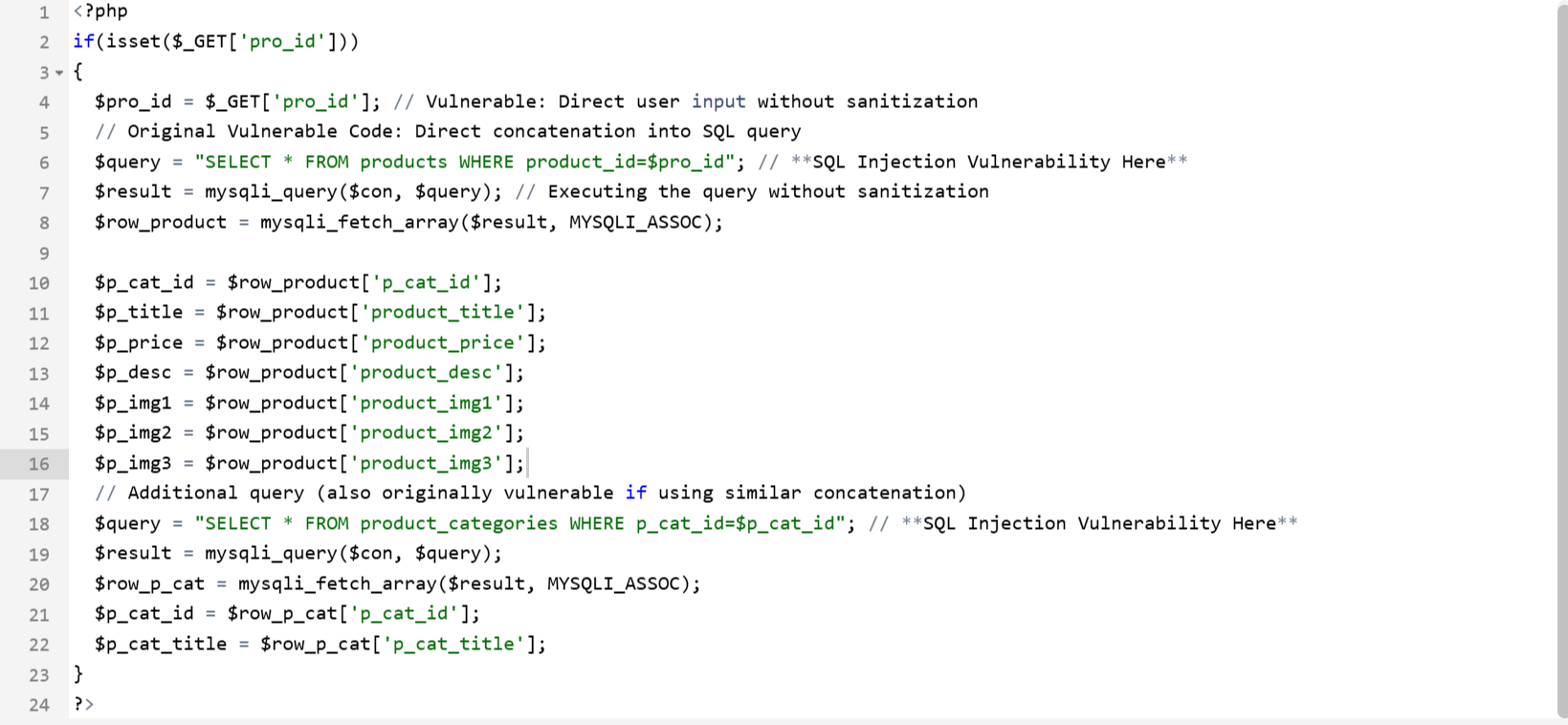}
    \caption{SQL Vulnerability Discovered on 'details.php' page.}
    \label{fig:figure4.19}
\end{figure}

\begin{figure}[h!]
    \centering
    \includegraphics[width=0.8\textwidth]{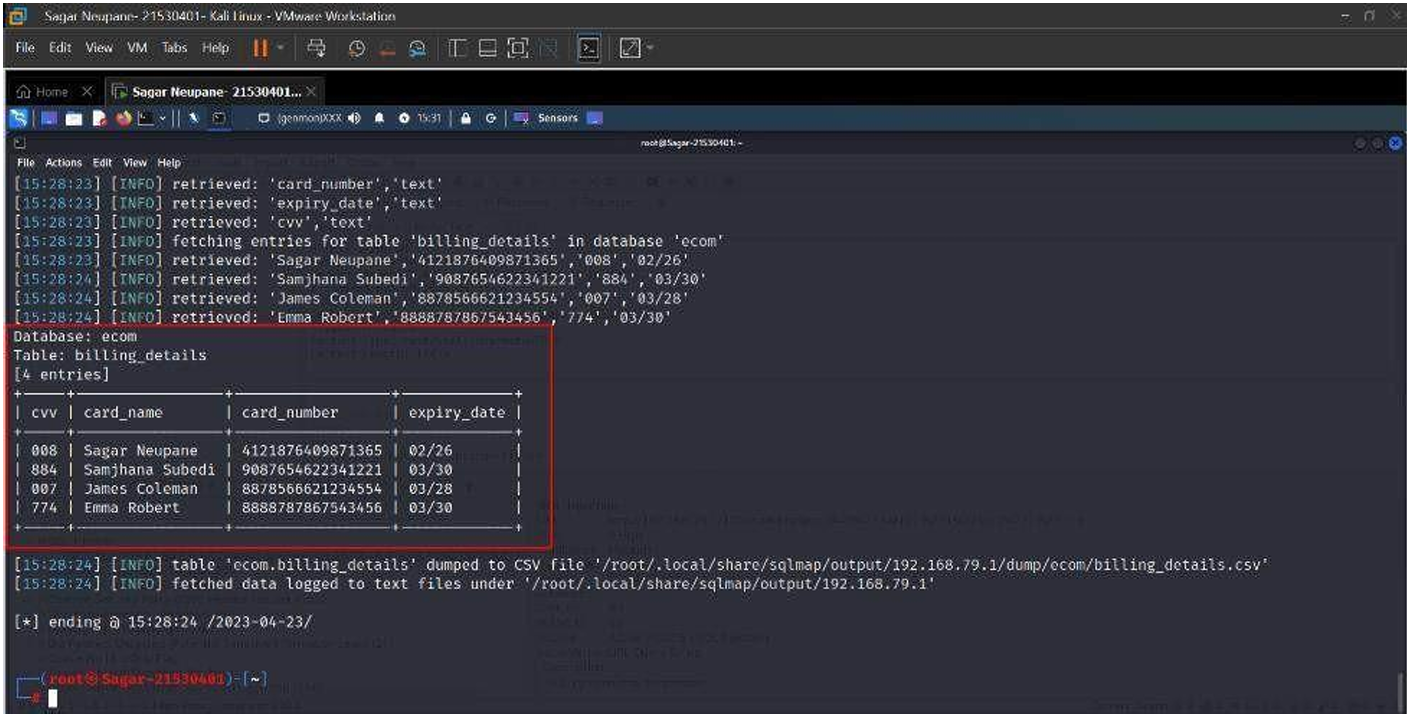}
    \caption{Dumping the `billing details` table.}
    \label{fig:figure4.30}
\end{figure}

\subsection{Patching}

The patching phase focused on remediating identified SQL injection vulnerabilities through secure code modifications, deployment, and rigorous verification to enhance website security. The vulnerability was traced to the detail.php page, where unsanitized GET parameters were directly concatenated into SQL queries. Remediation involved implementing prepared statements and parameter binding using PHP’s PDO library (Kromann, 2018), replacing vulnerable code with sanitized queries. Error handling was enhanced to prevent information leakage by incorporating custom error messages. The patched code was deployed to the XAMPP server, overwriting original files. Verification entailed rescanning the website with OWASP ZAP, which reported no SQL injection vulnerabilities, and re-running sqlmap with identical attack parameters, confirming the absence of exploitable flaws. The patching adhered to secure coding practices, including input validation and query parameterization, consistent with OWASP guidelines. This phase mitigated risks of data theft, confidentiality loss, and reputational harm.

\section{Results and Discussion}

This section presents the findings from the penetration testing, including identified vulnerabilities, exploitation outcomes, and the effectiveness of mitigation strategies.

\textbf{Data Theft}

Attackers may get access to and steal sensitive data from databases through SQL injection attacks. An e-commerce website that keeps consumer information, such as credit card information, is susceptible to SQL injection, for instance. A hacker who is successful in exploiting the flaw can retrieve credit card details from the database. Both the users who are impacted and the owner of the website could suffer considerable financial damages if this stolen data is utilised for financial theft or sold on the dark web. 

\textbf{Loss of Confidentiality}

By evading access rules and obtaining unauthorised data, SQL injection attacks might jeopardise the confidentiality of sensitive data. Think about a medical website that maintains patient medical records. A SQL injection vulnerability allows an attacker to get access to and retrieve private medical data, including diagnosis, treatments, and personal information. In addition to breaking privacy laws, this violation of confidentiality may also damage peoples' trust in the website and the healthcare provider.  

\textbf{Reputational Harm}

The penetration testing of a PHP and SQL-based e-commerce website revealed that SQL injection vulnerabilities significantly jeopardize organizational reputation by undermining user trust and fostering adverse public sentiment following a data breach. The testing process exposed exploitable weaknesses in the website’s product detail page, enabling unauthorized retrieval of sensitive information, including customer and admin credentials. This vulnerability could facilitate the public dissemination of compromised data, such as names, email addresses, and billing details, through illicit channels or online platforms, intensifying reputational damage. Such incidents erode stakeholder confidence, as customers and partners may view the organization as deficient in protecting critical data, prompting a shift toward competitors with robust security frameworks. The analysis underscores that reputational harm manifests as declining customer loyalty, reduced platform engagement, and lost business prospects. Moreover, negative media coverage and amplified social media discourse can prolong the reputational fallout, complicating efforts to restore brand integrity. The investigation further identifies indirect financial burdens, encompassing expenses for public relations campaigns, customer remediation, and enhanced security measures to rebuild trust. To counter these risks, the research advocates for proactive security strategies, including routine penetration testing and secure coding practices, such as parameterized queries and input sanitization, to eliminate vulnerabilities that precipitate breaches and their reputational repercussions. These findings resonate with established cybersecurity scholarship, which recognizes reputational harm as a pivotal consequence of data breaches, reinforcing the imperative for comprehensive defenses against SQL injection attacks in web applications.

\section{Mitigation Effectiveness}

The remediation process, detailed in the patching phase, utilized code sanitization and parameter binding to address vulnerabilities in the product detail page, where unsanitized GET parameters enabled malicious query injection. An updated vulnerability scan, conducted post-remediation using OWASP ZAP and sqlmap, confirmed the absence of SQL injection flaws, validating the effectiveness of these measures in preventing unauthorized database access. This outcome fulfills the objective of implementing and assessing secure coding practices to mitigate SQL injection risks. To sustain and enhance website security, several best practices are recommended, informed by the assessment’s findings and aligned with industry standards:

\begin{enumerate}
    \item \textbf{Input Validation}: Robust mechanisms should validate user inputs for type, length, and format before database interaction, employing filtering techniques to detect and block malicious content, thereby reducing the risk of injection-based attacks.
    \item \textbf{Parameterized Queries}: Prepared statements or parameterized queries (Downey and Fellows, 2012) should be adopted to separate SQL code from user inputs, automatically escaping inputs as data rather than executable code, effectively neutralizing SQL injection attempts.
    \item \textbf{Least Privilege Principle}: Database access should adhere to the principle of least privilege (Jero et al., 2021), granting users only the permissions necessary for their roles, minimizing potential damage from compromised accounts.
    \item \textbf{Regular Patching and Updates}: Software components, including frameworks and libraries, must be kept current with security patches to address known vulnerabilities that could be exploited for SQL injection attacks.
    \item \textbf{Web Application Firewall (WAF)}: Deploying a WAF provides an additional defensive layer, capable of identifying and blocking malicious SQL queries before they reach the database, enhancing protection against injection attempts (Prandl et al., 2015).
    \item \textbf{Security Testing and Code Reviews}: Routine penetration testing and code reviews should be conducted to identify and remediate vulnerabilities, ensuring ongoing detection of SQL injection and other security flaws.
    \item \textbf{Monitoring and Logging}: Comprehensive monitoring and logging systems should track anomalous activities, such as SQL injection attempts, enabling rapid detection and response to security incidents.
    \item \textbf{Security Incident Response Plan}: A well-defined incident response plan should outline procedures for containment, investigation, communication, and recovery in the event of a SQL injection breach, ensuring effective crisis management.
    \item \textbf{Continuous Security Assessment}: An ongoing commitment to security requires regular audits, threat monitoring, and adaptation of practices to address evolving risks, maintaining resilience against SQL injection threats.
    
\end{enumerate}
These strategies collectively address the identified vulnerabilities, which risked data theft, loss of confidentiality, and reputational harm, by establishing a (Hahn et al., 2015). The successful remediation and proposed recommendations underscore the importance of integrating technical, procedural, and strategic measures to achieve sustained mitigation effectiveness against SQL injection vulnerabilities in web applications.

\section{Limitations of this Research}

This research acknowledges several constraints that shaped its scope and generalizability, primarily due to the focused nature of the penetration testing. The assessment targeted a single, locally developed e-commerce website built with PHP and MySQL, hosted on a XAMPP server (Dvorski, 2007) in a controlled environment, which restricted the applicability of findings to diverse web applications employing other technologies, such as Python, Java, PostgreSQL, or Oracle. The isolated testing environment, devoid of real-world network conditions like variable latency or packet loss, limited the evaluation of vulnerabilities under operational scenarios. Additionally, the reliance on a limited toolset—OWASP ZAP (Nájera-Gutiérrez, 2016), sqlmap (Gunawan et al., 2018), and Nmap (Liao et al., 2020)—introduced potential biases, as alternative tools like Burp Suite (Wear, 2018) or Acunetix (Labiad et al., 2022) might have identified additional vulnerabilities or provided different insights into SQL injection risks.

Time and resource constraints curtailed the depth of manual testing, resulting in a heavy dependence on automated scans, which may have overlooked sophisticated attack vectors requiring advanced manual exploitation techniques. This research focused exclusively on SQL injection vulnerabilities, neglecting other prevalent web application threats, such as cross-site scripting (XSS) or cross-site request forgery (CSRF), which could interact with SQL injection risks in complex attack scenarios (Farah et al., 2016). The absence of real-world user interaction data hindered the simulation of dynamic attack patterns, as the website was tested in a static context without live traffic. Moreover, organizational and human factors, such as developer training or security policy implementation, critical for sustaining mitigation efforts, were not explored, underscoring the need for broader and more diverse testing frameworks to enhance the robustness of the findings.

\section{Future Work}

To address the identified limitations and advance web application security, this research proposes expanding the scope of penetration testing to encompass a wider range of web applications, incorporating diverse programming languages, frameworks, and database systems to enhance the generalizability of findings. Testing cloud-based applications and distributed architectures under real-world network conditions, such as latency and load balancing, is suggested to better simulate vulnerability exploitation scenarios. Integrating additional security tools, such as Burp Suite (Wear, 2018), Nessus (Kushe, 2017), or commercial scanners, is recommended to provide a more comprehensive vulnerability assessment, mitigating biases.

\section{Conclusion}

The application of automated tools like OWASP ZAP and sqlmap, complemented by manual testing, confirmed the presence of exploitable flaws, fulfilling the objective of conducting a thorough security assessment. The remediation process, involving code sanitization and parameter binding with PHP’s PDO library, effectively eliminated these vulnerabilities, as validated by subsequent scans, underscoring the efficacy of secure coding practices in preventing SQL injection attacks.

 The findings contribute to the broader field of web application security by providing practical insights into the detection and mitigation of SQL injection vulnerabilities, aligning with the research’s aim to enhance website security. The implementation of best practices, such as input validation, parameterized queries, and the least privilege principle, offers a robust framework for developers and security practitioners to safeguard web applications. These strategies address not only technical vulnerabilities but also organizational risks, such as reputational damage from data breaches, reinforcing the need for proactive security measures. By comparing the assessment’s outcomes with existing literature, this research validates the continued relevance of SQL injection as a prevalent threat and highlights the limitations of relying solely on automated tools, advocating for a balanced approach integrating manual expertise.

 The research’s practical application of penetration testing tools and secure coding techniques provides a valuable model for real-world security assessments, offering actionable recommendations for developers, security professionals, and organizations. By addressing SQL injection vulnerabilities, this research advances the discourse on web application security, advocating for continuous assessment and adaptation to evolving cyber threats to ensure resilient and trustworthy digital systems.

\bibliographystyle{ACM-Reference-Format}
\bibliography{reference}

\end{document}